# The university-industry knowledge relationship:

# Analyzing patents and the science base of technologies

Loet Leydesdorff [1]

University of Amsterdam, Science & Technology Dynamics

Amsterdam School of Communications Research (ASCoR)

Kloveniersburgwal 48, 1012 CX  Amsterdam, The Netherlands

loet@leydesdorff.net ; http://www.leydesdorff.net

**Abstract**

Via the Internet, information scientists can obtain cost-free access to large databases in the "hidden" or "deep web." These databases are often structured far more than the Internet domains themselves. The patent database of the U.S. Patent and Trade Office is used in this study to examine the science base of patents in terms of the literature references in these patents. University-based patents at the global level are compared with results when using the national economy of the Netherlands as a system of reference. Methods for accessing the on-line databases and for the visualization of the results are specified. The conclusion is that "biotechnology" has historically generated a model for theorizing about university-industry relations that cannot easily be generalized to other sectors and disciplines.

[1] I would like to thank Andrea Scharnhorst for comments on a previous draft of this paper.

## 1. Introduction

Perhaps even larger than the Internet itself are the resources which can be accessed and studied via the web. These resources are sometimes called the "hidden web," the "invisible web" or the "deep web" (Bergman, 2001; Sherman & Price, 2001). Unlike most web-based resources—which evolve and change with the development of the Internet during the years—some of the databases of the hidden web are certified and fixed. For example, the database of the U.S. Patent and Trade Office (USPTO) contains all U.S. patents since 1976 in html-format (at http://www.uspto.gov). The data have a legal status since a patent can be challenged in court, and therefore the text can no longer be changed after the patent has been issued. Furthermore, the examination by the patent examiner is highly codified (Granstrand, 1999).

The same data are offered on-line by commercial hosts like Dialog in formats that facilitate organization of the data and integrates it with data from other national or international (e.g., European) patent offices. The Derwent Innovation Index even offers an integration of the patent data with the *Web-of-Science* data of the Institute of Scientific Information. These commercial accesses, however, are relatively expensive for the purposes of academic research and higher education.

In this study, I explore the on-line data from an information theoretical perspective, that is, with a focus on how the knowledge base of the patents can perhaps be revealed. This question is theoretically interesting because patents have increasingly become a repository of information about how the socially organized production of scientific knowledge is interfaced with the economy (Noble, 1977). Organized knowledge production and control (Whitley, 1984) follows a logic of development and differentiation different from (potentially knowledge-based) innovation processes in the economy (e.g., Mansfield, 1989). The development of indicators for the knowledge base of an economic system can be considered a priority for innovation policies and the emerging program of innovation studies (e.g., David & Foray, 2002; Nelson, 1993; OECD/Eurostat, 1997; Leydesdorff & Meyer, 2003; Leydesdorff & Scharnhorst, 2003).

Patent data have been used extensively in economic geography, business economics, and macro-economics as indicators of the innovativeness of corporations, industries, and regions (Jaffe & Trajtenberg, 2002; Meyer, 2000; Pavitt, 1984). However, the specific interest of information scientists in how the patents relate to their knowledge base (e.g., Bhattacharya *et al*., 2003; Grupp & Schmoch, 1999; Narin & Olivastro, 1988, 1992) is not facilitated by using the value added by commercial databases like Derwent. The so-called "non-patent literature references" (NPLR) contain references to scientific journal literature and book chapters among other things, but this field has remained poorly organized in the commercial format. Abbreviations of journal names, for example, are not standardized.

In the case of scientific references, most patents provide titles between quotation marks in order to distinguish them from journal names or from the title of an edited volume. I will use this indicator as a point of access for exploring the knowledge base of patents. Because the practice of using quotation marks is almost exclusively the case for formalized literature,[1] I hypothesize that this indicator can be used as a proxy for accessing the knowledge base of patents.

[1] Sometimes newspaper articles are also included using this format.

Two domains will be explored for the year 2002:

1. All patents containing an address with the root "univ*" signifying "university" among the assignees.[2] Since 1980, the Bayh-Dole act in the United States and similar legislation in other countries granted universities the right to patents on the basis of federal funding. This led to an important increase in the participation of universities in the patenting domain (Henderson *et al.*, 1998; Sampat *et al.*, 2003). Universities can be among the assignees of patents. Inventor names remain natural persons.

2. For the comparison I have used the domain of all U.S. patents in 2002 with a Dutch address among the assignees or the inventors.[3] These patents can be considered as relevant to the knowledge base of the Netherlands as a national economy (Nelson, 1993).

**2. Methods and materials**

During 2002 a total of 184,531 patents were issued. 3,455 patents could be retrieved with the root "univ$" in the fields of addresses of inventors or assignees. After correction for words like "universal" and "universe," 3,291 patents remained that had been assigned to a university. As expected, the word "university" never figured among the inventor addresses. Note that a number of universities do not use the word "university" in their names (e.g.,

[2] The precise query was as follows: "isd/$/$/2002 and an/univ$". The $ is used as a wild card in the USPTO database, and therefore the query looks for all patent data that were issued in 2002 ("isd" = issue date) and that contain the root "univ" in the name of the assignee (field code: "an").

[3] The precise query was as follows: "isd/$/$/2002 and (acn/nl or icn/nl)". The abbreviation "nl" is used for the Netherlands; "acn" is the field code of the name of the country of the assignee and "icn" for the name of the country of the inventor.

MIT). Nevertheless, the delineation with the word "university" provides us with a convenient domain of patents for statistical exploration.

Second, patents with an origin in The Netherlands (as inventor or assignee) were downloaded as an example of a geographically contained set of foreign patent holders within the U.S. patent domain. In 2002 it happens to be the case that there are 1,963 patents with a Dutch assignee and equally 1,963 patents with a Dutch inventor. The combined set, however, contains 2,827 patents with a Dutch address (2,824 of these patents could be retrieved). More than national patents these foreign patents indicate an investment in the global marketplace.[4] The investments are made because of a value of the intellectual property to be protected.

The two domains are institutional and geographical, respectively. While the university-based patents can be used as an indicator of university-industry relations, the Dutch patents can be expected to represent the internationally oriented sectors of a national economy.

## 2.1 Methodological considerations

How does this difference in delineation influence the knowledge base of the corresponding sets of patents? This question will be pursued by analyzing the title words in the two sets and by relating these title words to the title words of the scientific documents cited in these sets. New developments in visualization software enable us to map asymmetrical matrices using large datasets.[5] Pajek, a freeware program for visualization developed by mathematicians at the University of Ljubljana,[6] for example, contains a subroutine for analyzing asymmetrical matrices both in either direction (Q- or R-mode structural analysis) or bimodal. Since innovations take place at interfaces, the mapping of asymmetries at interfaces in terms of variation, selection, and codification can be considered a priority from the perspective of evolutionary economics and innovation studies (David & Foray, 2002; Leydesdorff, 2003).

[4] In 2002, the number of patents with a Dutch address among the assignees in the database of the European Patent Office is 3,193 and the number of patents with a Dutch address among the inventors is 2,667. The combined set (with an OR) is 3680. Thus, the number of patents published in the USPTO database is 76.8% of the number of patents published with a Dutch address in the EPO database. Note that these two sets do not have to be based on the same patents.

Both the analysis of title words and the analysis of the interfaces with the NPLR will be visualized using the algorithm of Kamada & Kawai (1989) as it is available in Pajek. This algorithm represents the network as a system of springs with a relaxed lengths proportional to the edge length. Nodes are iteratively repositioned to minimize the overall "energy" of the spring system using a steepest descent procedure. The procedure is analogous to some forms of non-metric multi-dimensional scaling.[7] I will compare the results of this relational analysis with the results of a (positional) factor analysis of the same map (Burt, 1982; Leydesdorff, 1995). In order to keep the visualizations readable, the analysis will pragmatically be limited to the approximately one hundred most frequently occurring words for each case.

As a similarity measure among vectors of word distributions in titles of patents I shall use the cosine (Salton & McGill, 1983). This measure has an advantage over the Pearson correlation (used by the factor analysis) in that the similarity is insensitive to the number of zeros because the cosine is not based on the mean of the distribution (Larsen & Ingwersen, 2002;

---

[5] Limited datasets could previously be mapped asymmetrically using Quasi-Correspondence Analysis (Tijssen *et al*., 1987).

[6] The homepage of Pajek can be found at http://vlado.fmf.uni-lj.si/pub/networks/pajek

[7] A disadvantage of this model is that unconnected nodes may remain randomly positioned across the visualization. Unconnected nodes are therefore not included in the visualizations below. See for more details about the different algorithms, for example, the overview in the introduction to the social network image animator software package SoNIA at http://www.stanford.edu/~skyebend .

Ortega Priego, 2003; cf. Ahlgren *et al.*, 2003; White, 2003).[8] In empirical cases, these two similarity measures lead often to similar results (Leydesdorff & Zaal, 1989).[9] In the case of the asymmetrical (bi-modal) matrices of words in the titles of patents versus words in the titles of the corresponding literature references, the cell values are not normalized.

### 2.2. The retrieval of data from the web

Patent data are brought on-line by the U.S. Patent and Trade Office (at http://www.uspto.gov) and by the European Patent Office (at http://ep.espacenet.com). The latter database also contains the data of the World Patent Organization. However, the European and world patents are not fully standardized and partly in other formats, while the U.S. database is standardized, organized in hypertext mark-up language (html), and accessible for searching by robots.[10] Furthermore, the U.S. database is often used in scientometric research for comparative purposes because it standardizes the presence of other nations in a single representation (Narin & Olivastro, 1988, 1992). This database allows, among other things, for the retrieval of citation patterns in terms of both the previous patents cited and the scientific (that is, non-patent) literature cited. Additionally, the follow-up in terms of 'being cited' in later patents can be traced.

---

[8] Salton's cosine is defined as the cosine of the angle enclosed between two vectors *x* and *y* as follows:

$$\text{Cosine}(x,y) = \frac{\sum_{i=1}^{n} x_i y_i}{\sqrt{\sum_{i=1}^{n} x_i^2}\sqrt{\sum_{i=1}^{n} y_i^2}} = \frac{\sum_{i=1}^{n} x_i y_i}{\sqrt{(\sum_{i=1}^{n} x_i^2)*(\sum_{i=1}^{n} y_i^2)}}$$

[9] The Jaccard Index differs not only with a factor two from the cosine (Hamers *et al.*, 1989), but leads in empirical cases often to results which are rather different from the Pearson correlation (Leydesdorff & Zaal, 1988). Strong relations in the database (segments) are fore-grounded by the Jaccard Index, while Salton's cosine organizes the relations geometrically so that they can be visualized as structural patterns of relations (Luukkonen *et al.*, 1993; cf. Michelet, 1988). Factor or eigenvector analysis enables us to analyze this construct in terms of its orthogonal dimensions (Wagner & Leydesdorff, 1993).

[10] The USPTO states a limitation on bulk downloads of the data at http://www.uspto.gov/patft/help/notices.htm.

Because of its legal status within the American administration, the USPTO database is extremely well organized in terms of search terms and reliability. The search options are documented with help screens (Black, 2002). The results can be retrieved with screens of fifty titles consecutively. These titles are hyperlinked with the full texts of the patents containing all the information available in html format. The labels are consistent and therefore the data can conveniently be parsed and brought under the control of relational database management.

Web search engines do not go more than two levels deep into the USPTO's Web site because they cannot query a database.[11] However, the on-line retrieval can be automated by using a routine in Visual Basic. Visual Basic 6 was the first version to contain a so-called Internet Transfer Control. This component enables us to download the data from a structured database at the Internet such as the one under study here. The routine for searching all the patents in 2002 with a Dutch address among the inventors or assignees is provided as an example in an Appendix 1. By cutting and pasting the search command from the Internet search in the long line which defines the URL string as the variable named "strURL," one is able to accommodate this routine to one's specific requirements. The parameter N controls the record number and the parameter P reflects that the titles are provided in screens of fifty records consecutively. The download, parsing, and organization into a relational database management can thus be fully automated. This allows researchers to exploit this data without constraints.[10]

---

[11] Search engines do not access URLs which contain a question mark because this indicates the use of script technology. If spiders encounter a "?" in an URL or link, they are programmed to stop crawling because they could encounter poorly written script or intentional "spider traps" (at http://www.lib.berkeley.edu/TeachingLib/Guides/Internet/InvisibleWeb.html#Why2; cf. Reddi *et al.*, 2003)

As the reader will note, a similar routine can be written for any database which provides a systematic indication of the sequential results (e.g., the *AltaVista Advanced Search Engine*). However, some databases have deliberately blocked this mode of searching by a robot (e.g., ISI's *Web of Science*, *Google*).[12]

## 3. Results

### *3.1 University-based patents*

Among the 184,531 patents with an issue date in 2002, 3,291 refer to universities among the names of the assignees. The total number of assignees is 3,823 and the total number of inventors 9,217. In sum, not many of these patents are co-assigned, but many of them are co-invented. The 3,291 records contain 44,268 references to patents and 62,138 references to non-patent literature. The number of scientific references outnumbers the patent references for this university-based sample.

These patents contain 5,148 unique words (after correction for the stopwords).[13] The 102 most frequently occurring words among these are used for the visualizations in Figures 1 and 2 below. The words included occur with a frequency of more than 26 times.

[12] The search engine Google offers an alternative by using its own so-called APIs. These allow for searching the database also on dates, albeit using the Julian calendar. The *AltaVista Advanced Search Engine* is hitherto the only database allowing for searching with calendar dates (Leydesdorff, 2001).
[13] For reasons of consistency, the stopword list available at http://www.uspto.gov/patft/help/stopword.htm was used throughout this study as a standard corrective to the inclusion and exclusion of common words. Otherwise, the words are corrected only for the plural "s."

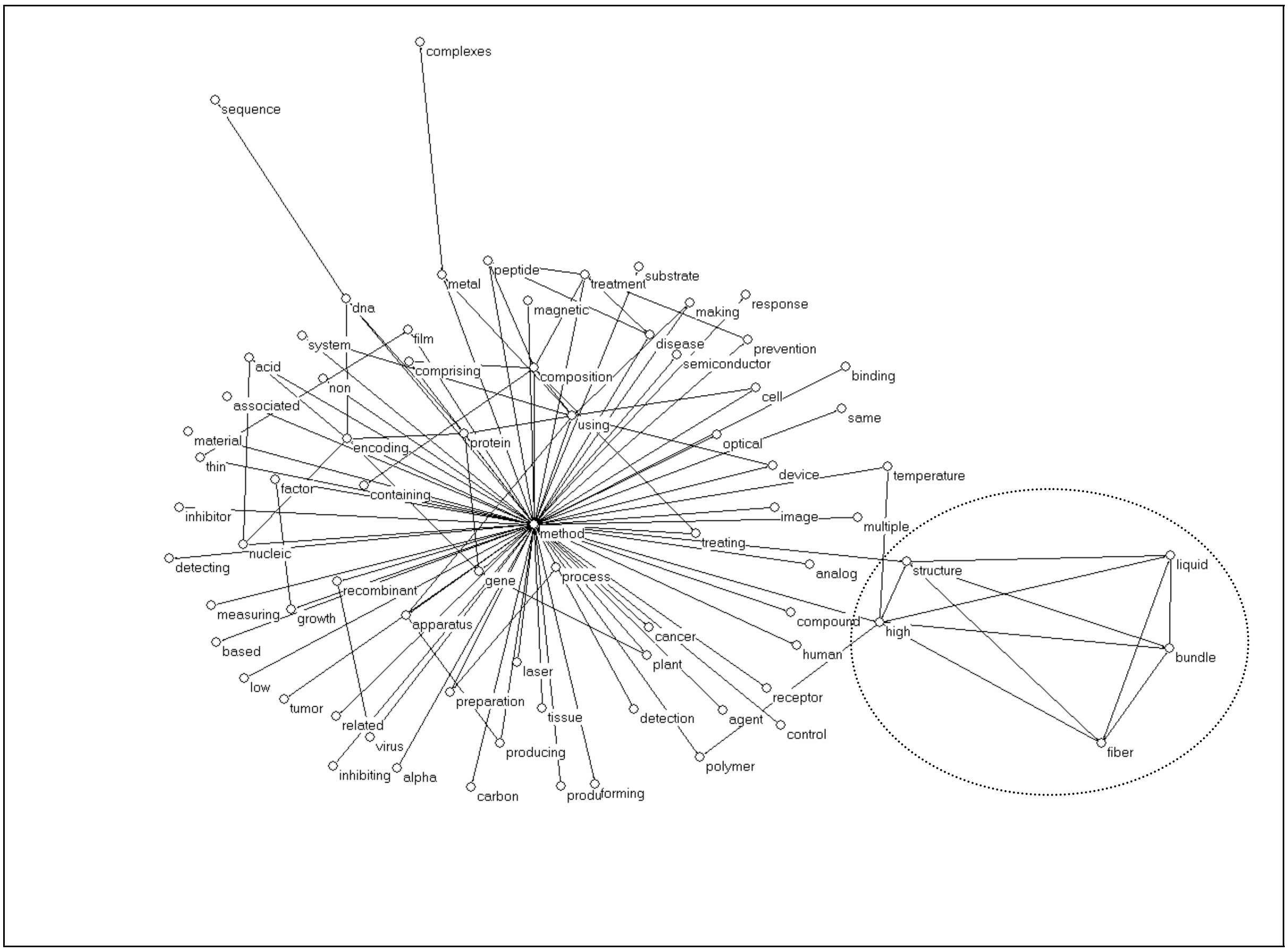

**Figure 1**

Co-occurrence network of 102 title words in patents with a university address during 2002 (N Patents = 3291; Word frequency > 26; 75 words connected at the threshold level of co-occurrences ≥ 10).

Figure 1 shows the co-word map given a threshold of ten co-occurrences and before the normalization. It is clear that the most frequently occurring word is “method(s).” This word draws most of the other words into a star-shaped network. However, one cluster is visible containing the words “fiber,” “liquid,” “bundle,” etc. This cluster is related to the central set through the words “polymer,” “structure,” “high,” and “temperature.”

Upon normalization using the cosine formula,[5] the picture changes to exhibit the intellectual organization (Figure 2). The main areas of technological activities in which universities patent are now visible as clusters. The relatively low value of the threshold (that is, cosine ≥ 0.1) indicates that this structure is relatively robust.

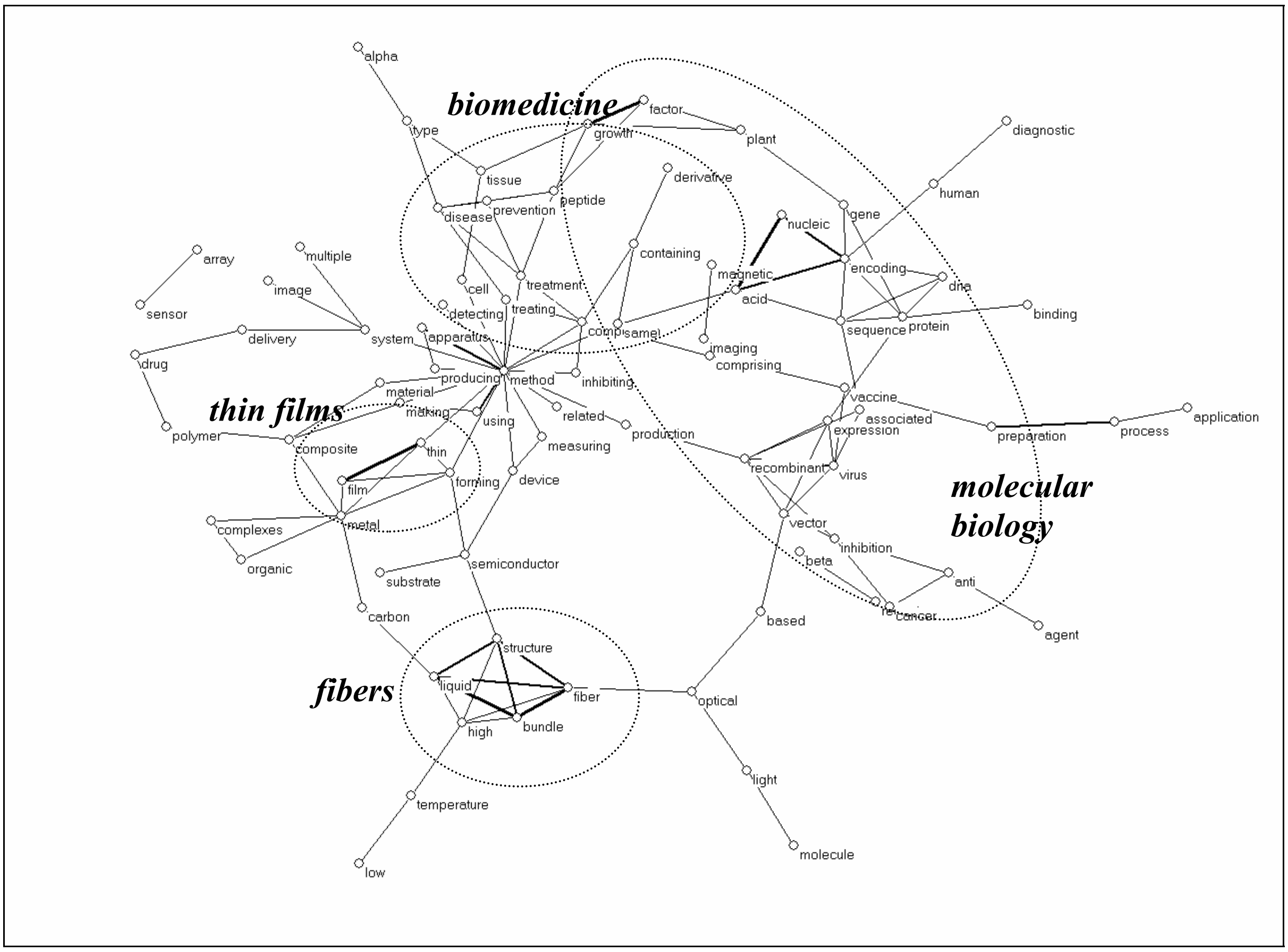

**Figure 2**

Cosine normalized map of 102 title words in 3,291 patents with a university address in 2002. (N Patents = 3,291; Word frequency > 26; 85 words connected at the threshold level of cosine ≥ 0.1).

The various clusters indicated in the map can be further explored and designated using factor analysis of the matrix. The factor analysis positions the clusters differently in a multi-dimensional space. Figure 3 shows the plot of components 1 and 2 in a six-factor solution of this matrix. (Note that factor analysis uses by default the Pearson correlation as a similarity measure.)

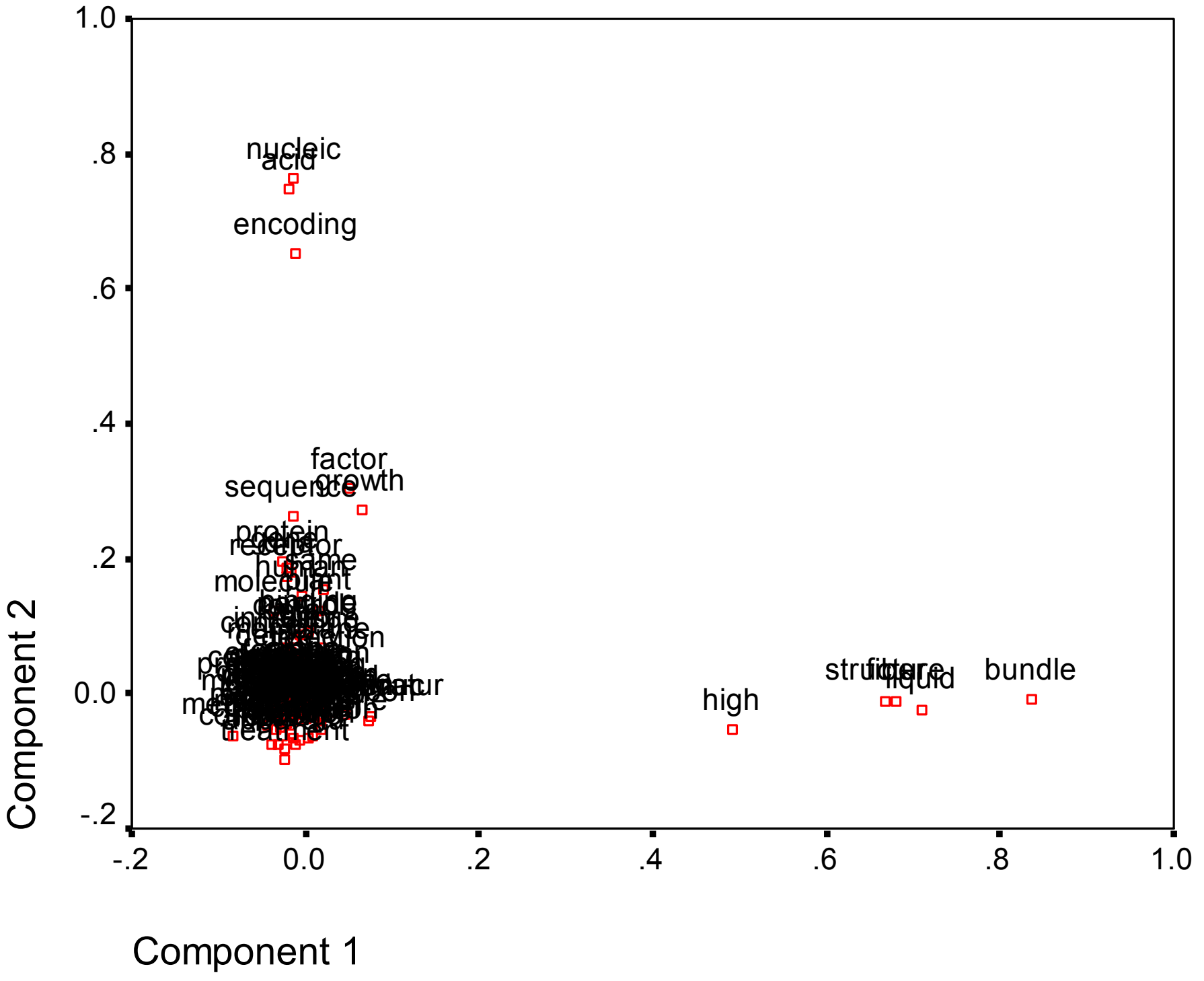

**Figure 3**

Results of the factor analysis of the co-occurrences of 102 title words in 3,291 patents with a university address in 2002.

Figure 3 exhibits two of the clusters indicated in Figure 2 as the major dimensions of the matrix. Another (fifth) factor (with factor loadings for the title words “growth,” “sequence,” and “factor”) exhibits interfactorial complexity with the one designated above as “molecular biology” (containing the title words “nucleic”, “acid”, and “encoding”). In principle, one would thus be able to use the factor loadings of words in a positional analysis for purposes of the designation of clusters in the visualization of mutual relations (Burt, 1982). However, the development of this technique would reach beyond the scope of this study (Leydesdorff, in preparation).

*3.2 The scientific knowledge base of university-based patents*

The 38,509 literature references that contain title words within quotation marks can be broken down into 25,078 unique words. Of these words, we use again the approximately one hundred words which occur most frequently. These one hundred words were found to occur more than 438 times in these references. The co-occurrences of these words with the 102 title words of patents used in the previous analysis can be organized into an asymmetrical or bi-modal matrix (Table 2). The title words of the patents are used as the column variables and the title words in the NPLRs as the case labels. Using Pajek this matrix can be represented as in Figure 4.

102 title words of patents →

| | *method* | *system* | *using* | *apparatus* | *composition* | *protein* | *cell* |
|---|---|---|---|---|---|---|---|
| *cell* | 2087 | 182 | 414 | 26 | 653 | 466 | 1210 |
| *protein* | 1549 | 121 | 374 | 34 | 501 | 956 | 412 |
| *gene* | 1788 | 275 | 329 | 31 | 678 | 544 | 772 |
| *human* | 1150 | 134 | 211 | 16 | 477 | 230 | 535 |
| *dna* | 706 | 116 | 84 | 8 | 190 | 156 | 254 |
| *expression* | 640 | 94 | 102 | 2 | 229 | 181 | 372 |
| *receptor* | 683 | 3 | 91 | 2 | 198 | 197 | 175 |
| *virus* | 686 | 198 | 184 | 2 | 187 | 212 | 803 |
| *factor* | 525 | 16 | 102 | 5 | 178 | 81 | 174 |
| *tumor* | 570 | 42 | 121 | 2 | 322 | 138 | 210 |
| *synthesis* | 438 | 36 | 121 | 13 | 160 | 79 | 84 |
| *peptide* | 586 | 27 | 180 | 11 | 177 | 297 | 141 |
| *growth* | 498 | 22 | 165 | 10 | 164 | 56 | 207 |

100 title words in references

↓

**Table 2**

Part of the asymmetrical matrix of 102 title words in patents versus 100 most frequently occurring title words in the literature references within these patents.

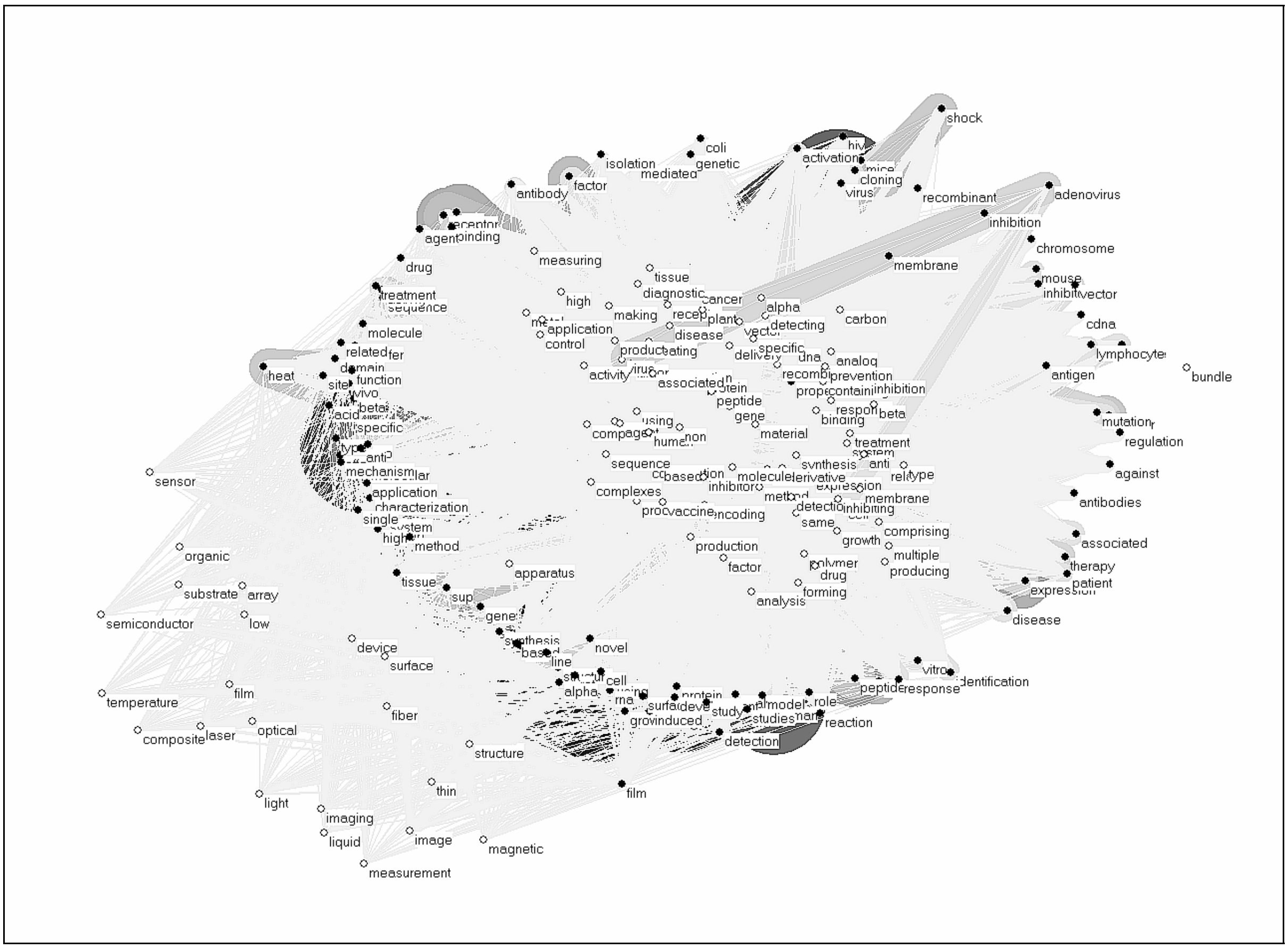

**Figure 4**

Bimodal representation of the 102 most frequently occurring words in titles of patents (> 26 times) and the 100 most frequently occurring words in titles of scientific citations (> 438 times) in these patents (3291 patents; 38,509 scientific references).

The figure shows that the title words of the patents in the biomedical sector are sorted in the middle of a set of title words from the cited scientific literature. The connecting patent words (in white) are in the center because this set ties the set of words from the literature together with various foci. A group of patent words on the bottom-left side of this figure is only related to the research side of the biomedical literature and not to more clinically oriented words like “patient,” “therapy” or “treatment.” Title words from these patents are positioned outside the halo formed by the title words of the NPLR.

The depiction in Figure 4 should be read with the caveat that it remains a visualization of a system. The system under study may have more than a single (relative) minimum for the "energy" (Kamada & Kawai, 1989). Another local minimum, for example, can be found in this matrix so that the title words of patents envelop the title words of the NPLR. The patents draw on the scientific references as their knowledge base and this can be depicted either as a focusing device in the middle of the variation or as a ball with the original variation contained within it.

If we focus on the subset of 1,920 patents which contain NLPR, we can generate a list of words occurring most frequently in these patents only. The 96 words occurring more than 17 times in these patents are mapped in Figure 5 after normalization for cosine $\geq$ 0.1 (as previously). As can be expected on the basis of the above analysis, the biotechnology field and the molecular biology field are completely dominant in this subset. Note that the agricultural applications of biotechnology are marginal.

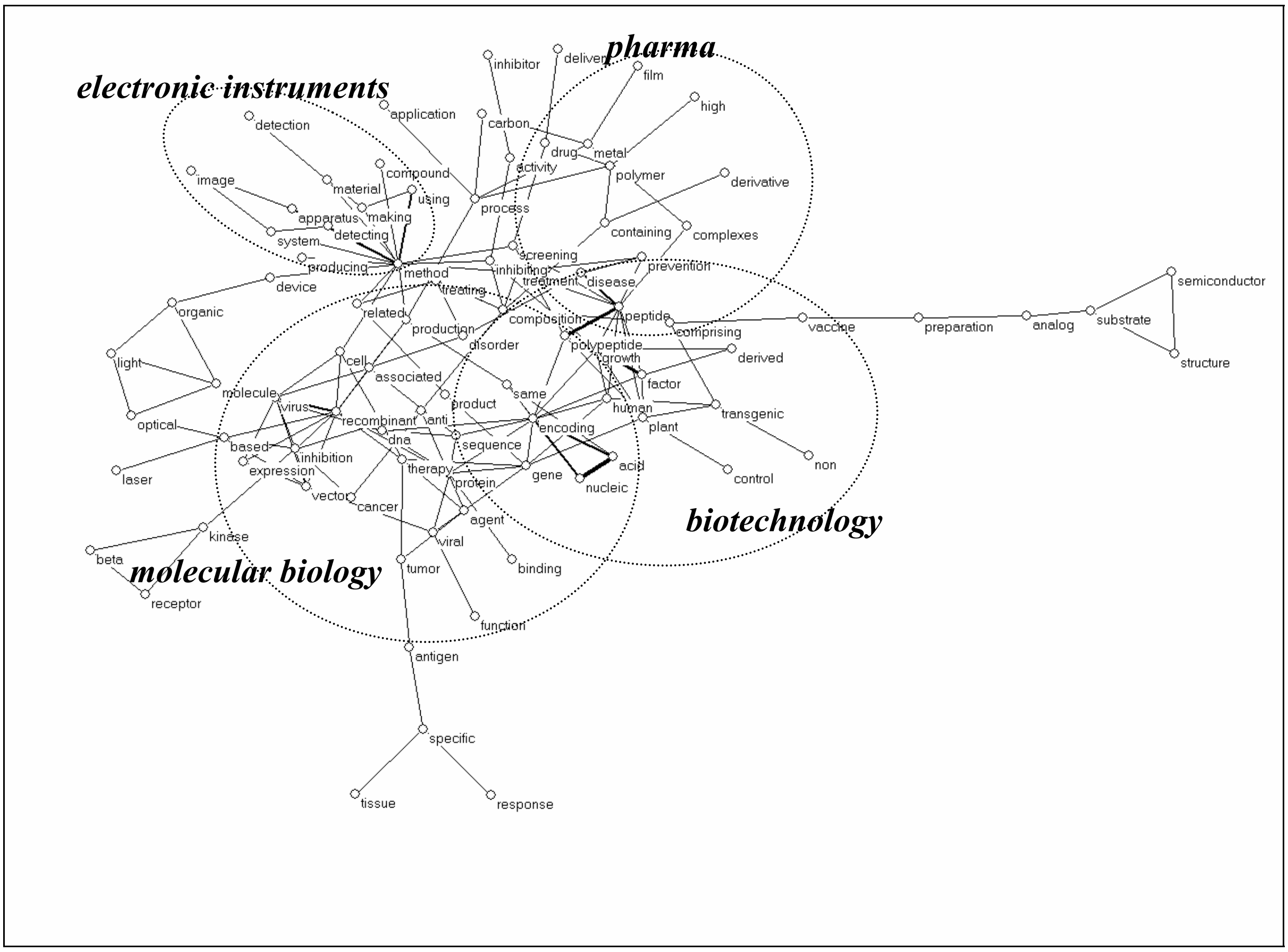

**Figure 5**

Cosine normalized map of relations between 96 title words in patents with literature references and a university address in 2002. (N Patents = 1,920; Word frequency > 17; 90 words connected at the threshold level of cosine $\geq$ 0.1).

*3.3 Dutch patents*

Unlike the patents with a university address among the assignees, the 2,824 patents with a geographical address in the Netherlands contain far more references to other patents than to non-patent literature, notably 31,514 and 6,396 references, respectively. Among the 6,396 non-patent literature references some 3,440 contain title words between quotation marks, and these reference belong to only 643 patents in the set. Thus, the science base of this set—as indicated by formal literature references—is much less prominent than in the case of the previous set. The role of the Dutch universities is marginal: while 29 of these 2,824 patents contain a university address, only 15 of these university addresses (0.5%) are located in the Netherlands.

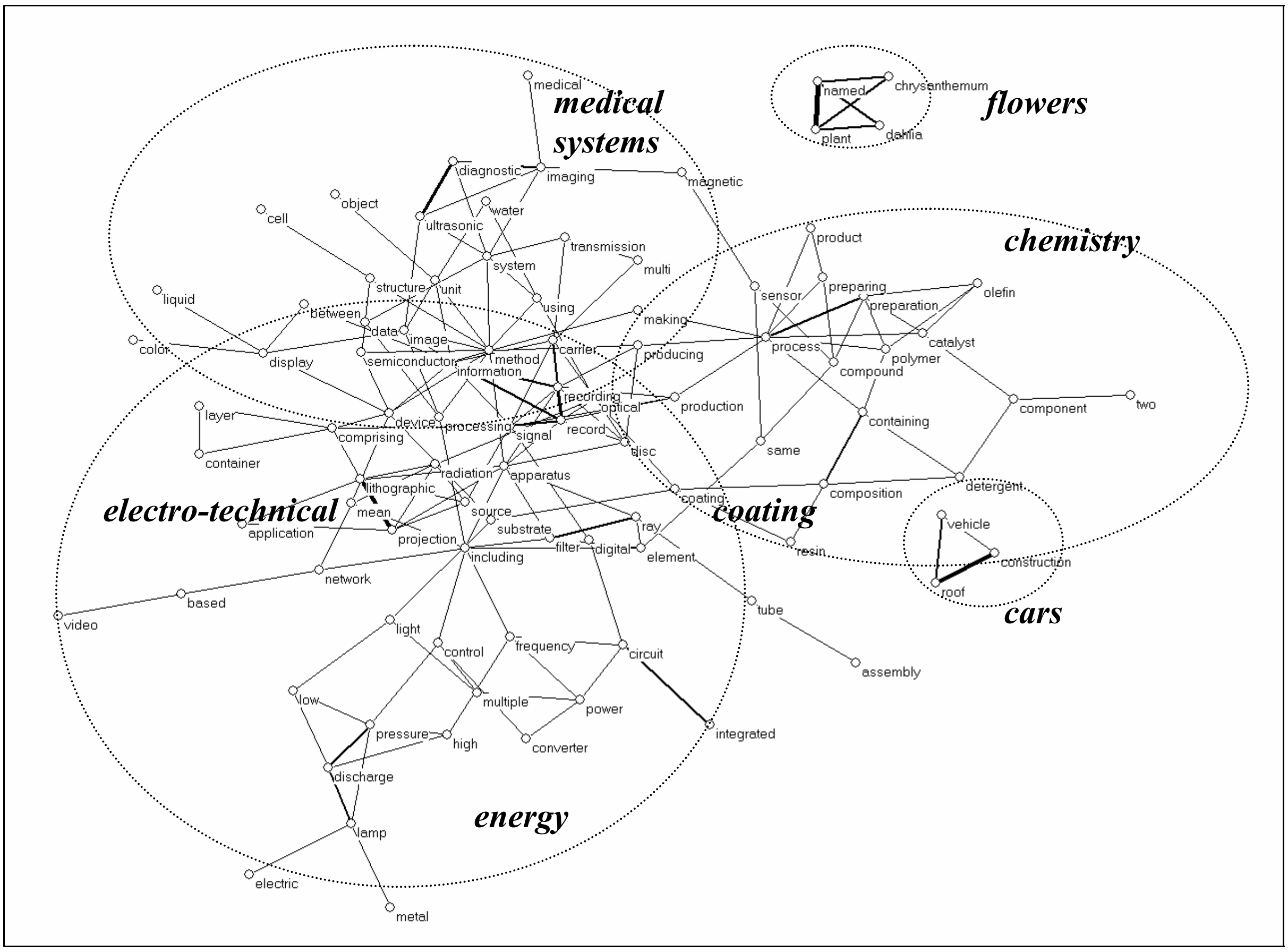

**Figure 6**

Cosine normalized map of 105 co-occurring words in patents (in 2002) with a Dutch address among the assignees or inventors (N Patents = 2,824; Word frequency > 22; 94 words connected at the threshold level of cosine ≥ 0.1).

Figure 6 shows the normalized co-occurrence map of the 105 title words that occur with a frequency of more than 22 among the 4,005 unique title words contained in the set. The picture exhibits a recognizable representation of the Dutch industrial structure with a dominance of electro-technical and chemical applications. Multinational corporations are dominant in the set. For example, Philips with a focus on electro-technical systems holds 768 of the 1,963 patents (39.1%) with a Dutch address among the assignees. Medical systems are related to the electro-technical side of the set through imaging devices. The occurrence of a small set of patents related to the names of flowers is noteworthy.

Figure 7 exhibits the occurrence of these 105 title words of patents in relations to the 3,440 scientific literature citations that contain title words between quotation marks. The latter constitute a domain of 6,072 unique words, of which we selected the 101 that occur with a frequency larger than 31.

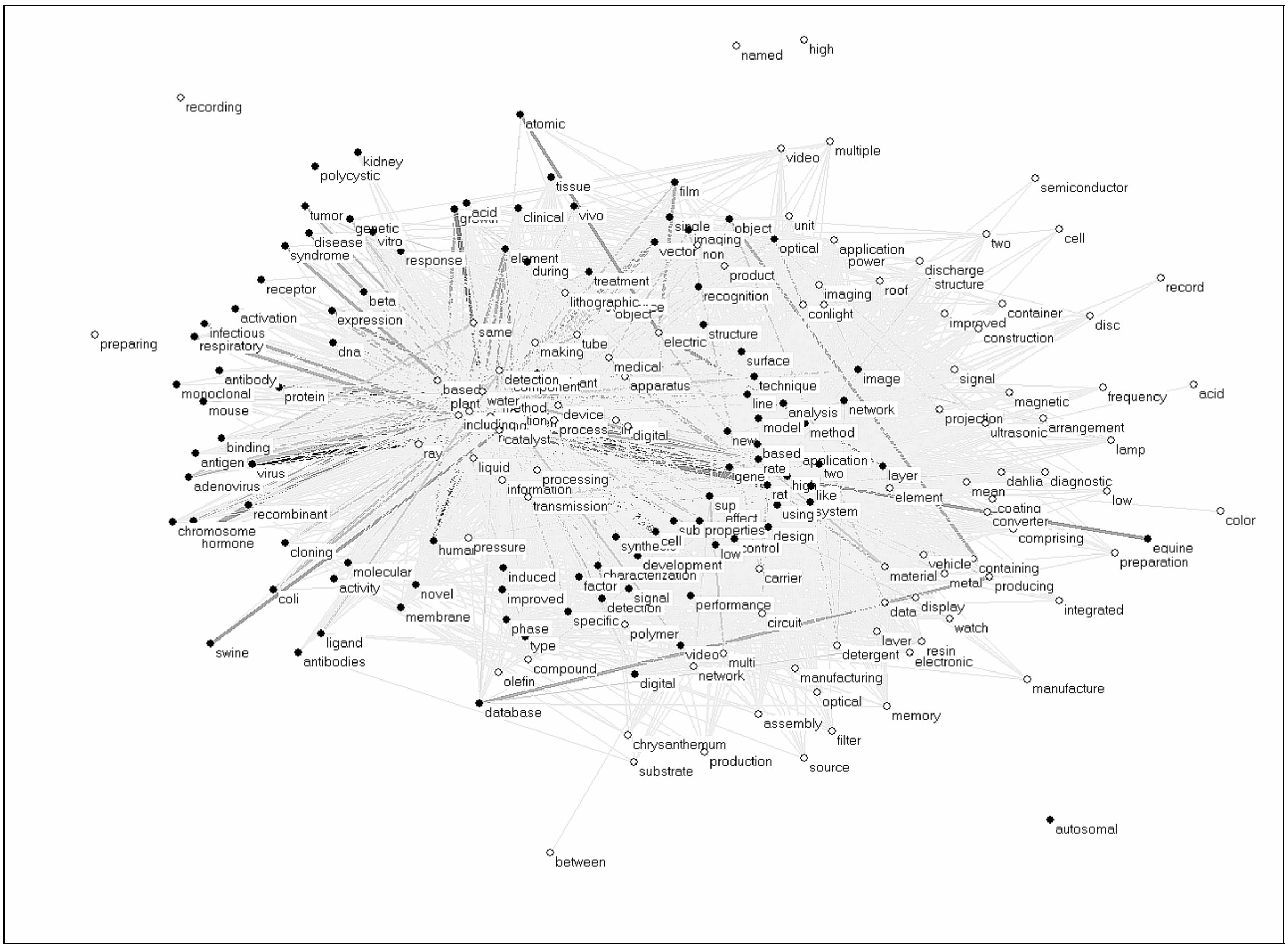

**Figure 7**

Network of title words in patents[14] with a Dutch address among the assignees or inventors in relation to the title words used in their literature references (N Patents = 2,824; Word frequency > 22; 3,440 literature references with 6,072 unique words of which 101 occur with a frequency > 31).

The picture shows that the references are concentrated in the bio-medical sector. A relatively small set of (643) patents is related in this way, and these title words are placed in the center of a halo with the title words from the literature references. The other patent words are not related in this way. Title words of these patent are placed outside the halo of bio-medical applications.

[14] Because I used the first ten characters of the words for the identification in this case, the 105 words were reduced to 104: "manufacture" and "manufacturing" were equated.

The 643 “science-based” patents contain 1,681 unique words of which 107 occur more than 6 times. Figure 8, finally, provides the cosine-map of these co-occurrences in a format similar to the ones above. Note that this map is much more fine-grained than the previous ones because of the much lower level of the threshold for the occurrence frequencies of title words. Because of the smaller range of values in the cells, the structure becomes visible only when the cosine-threshold is raised to ≥ 0.2.

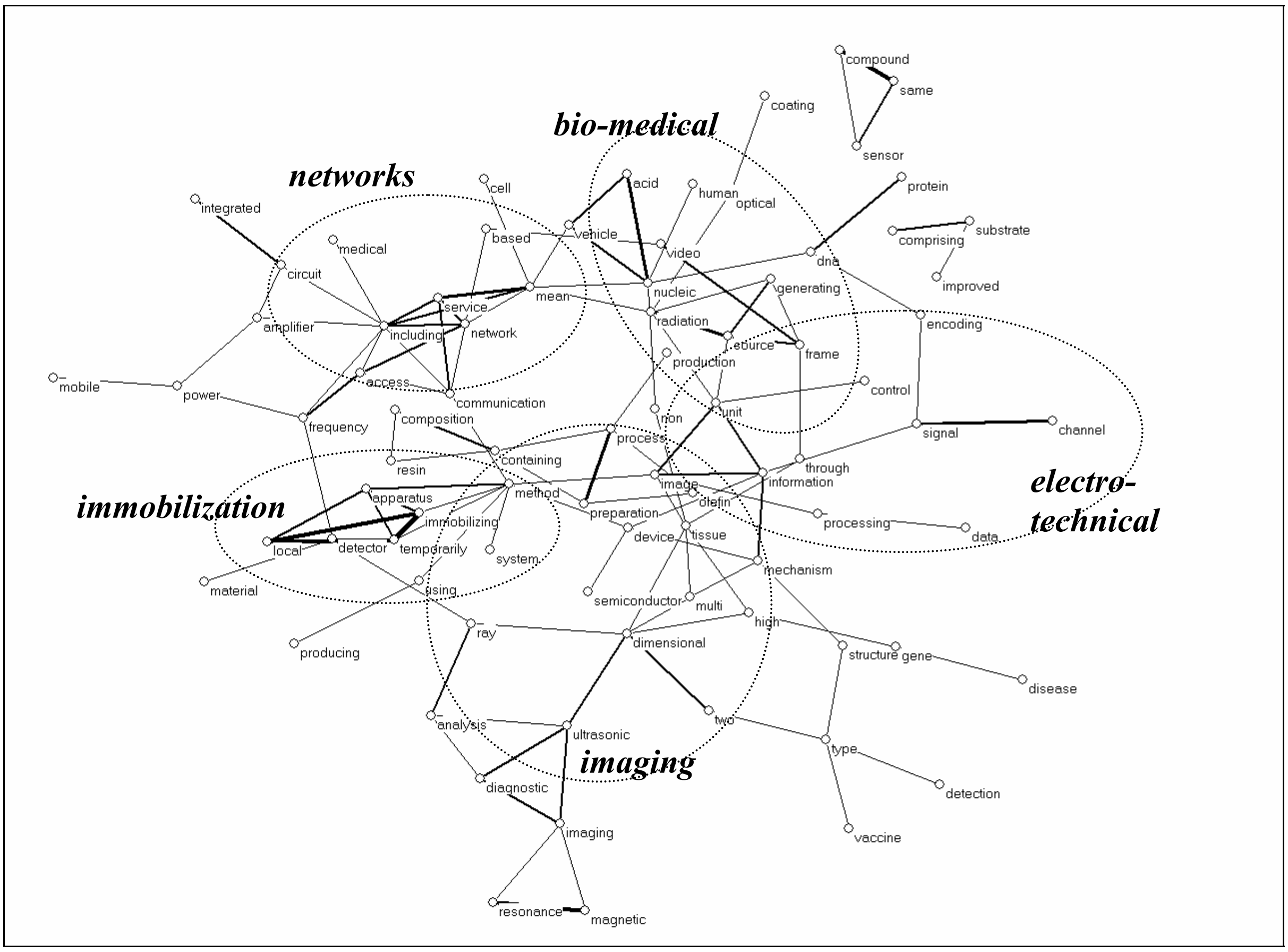

**Figure 8**

Cosine normalize map of 107 most frequently occurring words in 643 “literature-based” patents with a Dutch address among the assignees or inventors (N Patents = 643; Word frequency > 6; 83 words connected at the threshold level of cosine ≥ 0.2).

Figure 7 above showed that the knowledge base of the Dutch patents—as visible in the U.S. patent database—is integrated by the bio-medical applications, but Figure 8 shows that the latter are not central to the aggregate of these activities. In this depiction, the industrial structure remains more important than the intellectual organization of these patents. Biomedical terms (e.g., "DNA", "nucleic") are relatively peripheral in Figure 8. However, the finding that the *knowledge base* of this patent set is integrated by a bio-medical network of title words in their NPLR is meaningful because the industrial structure visible at the surface is dominated by electro-technical and chemical applications.

## 4. Conclusion

The science-based model of university-industry collaborations was shaped in the 1980s with biotechnology as the prime example (Narin & Noma, 1985; OECD, 1988). Our data for 2002 suggest that this pattern has now been established as a dominant pattern (Narin *et al.*, 1997; Owen-Smith *et al.*, 2002). Information and communication technologies, for example, have not led to similar patterns of formalized exchanges between the scientific literature and the patent literature in other fields. Kaghan & Barnett (1997) signaled that the laboratory model tends to work as a "metonymy" because it guides the thinking about new policies in university-industry relations. Universities are active in new fields (e.g., thin films; cf. Bhattacharya *et al.*, 2003), but the relationship with the organized knowledge production system is much less formalized in terms of literature relations.[15]

[15] Glänzel & Meyer (2003) have studied the "reverse citation" of patents being cited in the scientific literature. Their conclusion is that established fields like "chemistry" are the main contributors to these relations.

In the second case, we turned to The Netherlands as an example of a knowledge-based, but nationally integrated economy. When we raise the question of how this knowledge base was reflected in the U.S. patent data, one can recognize the major industrial players in the patent domain. As noted, the role of the Dutch universities is marginal. Nevertheless, among these patents, only the ones with bio-medical relevance contain the noted pattern of science-based references. Thus, this relationship between scientific literature and patents is not specific to universities, but sector specific. In the Dutch case, the patterns of scientific referencing provide a network that connects the main operational areas of knowledge-based industries in the background.

These results suggest that one should be aware that policy-makers tend to think about university-industry relations in general terminologies, but that these relations are mainly shaped in the knowledge base of the bio-medical sector. Other sectors may contain mechanisms for integration and knowledge-transfer that are completely different from these bio-medical innovations.  Thus, one should not generalize easily from the experience with biotechnology and bio-medicine to other sectors of industry or disciplines of science. Biotechnology is a specific mode of interrelationship between science and industry.

From the perspective of the further development of Internet research and information science and technology, I have mainly wished to show that the Internet opens domains beyond the Internet for new scientific investigations. One routine for accessing these "hidden" domains was specified. The large amounts of data that can be made available by this technique, can be analyzed by using the visualization tools and the normalizations indicated.

## References

Ahlgren, P., B. Jarneving, and R. Rousseau (2003). Requirement for a Cocitation Similarity Measure, with Special Reference to Pearson's Correlation Coefficient, *Journal of the American Society for Information Science and Technology* 54(6), 550-560.

Bergman, M. K. (2001). The Deep Web: Surfacing Hidden Value, *Journal of Electronic Publishing,* 7(1) (2001); at http://www.press.umich.edu/jep/07-01/bergman.html

Bhattacharya, S., H. Kretschmer, and M. Meyer. (2003). Characterizing Intellectual Spaces between Science and Technology. *Scientometrics* 58(2), 369-390.

Black, G. R. (2002). *Exploring Inventions and Ideas: Keyword Patent Searching - Online*. Southfield, MI: http://www.keypatent.net/ .

Burt, R. S. (1982). *Toward a Structural Theory of Action*. New York, etc.: Academic Press.

David, P. A., & D. Foray. (2002). An Introduction to the Economy of the Knowledge Society. *International Social Science Journal*, 54 (171), 9-23.

Fuchterman, T., and E. Reingold (1991). Graph drawing by force-directed replacement, *Software--Practice Experience* 21, 1129-1166.

Glänzel, W., & M. Meyer. (2003). Patents Cited in the Scientific Literature: An Exploratory Study of 'Reverse' Citation Relations. *Scientometrics* (forthcoming).

Granstrand, O. (1999). *The Economics and Management of Intellectual Property: Towards Intellectual Capitalism*. Cheltenham, UK: Edward Elgar.

Grupp, H., and U. Schmoch. (1999). Patent Statistics in the Age of Globalisation: New Legal Procedures, New Analytical Methods, New Economic Interpretation. *Research Policy, 28*, 377-396.

Hamers, L., Y. Hemeryck, G. Herweyers, M. Janssen, H. Keters, R. Rousseau, et al. (1989). Similarity Measures in Scientometric Research: The Jaccard Index Versus Salton's Cosine Formula. *Information Processing & Management*, 25 (3), 315-318.

Henderson, R., A. Jaffe, and M. Trajtenberg (1998). Universities as a Source of Commercial Technology: A Detailed Analysis of University Patenting, 1965-1988. *Review of Economics and Statistics*, 80(1), 119-127.

Jaffe, A. B., & M. Trajtenberg. (2002). *Patents, Citations, and Innovations: A Window on the Knowledge Economy*. Cambridge, MA/London: MIT Press.

Kaghan, William N. and Gerald B. Barnett (1997). The Desktop Model of Innovation in Digital Media, in: Henry Etzkowitz and Loet Leydesdorff (eds.), *Universities and the Global Knowledge Economy: A Triple Helix of University-Industry-Government Relations*. London: Cassell Academic, pp. 71-81.

Kamada, T., and S. Kawai (1989). An algorithm for drawing general undirected graphs, *Information Processing Letters* 31(1), 7-15.

Larsen, B., & P. Ingwersen. (2002, August 11-15). The Boomerang Effect: Retrieving Scientific Documents Via the Network of References and Citations. Paper presented at *SIGIR'02*, August 11-15, Tampere, Finland.

Leydesdorff, L. (1995). *The Challenge of Scientometrics: The Development, Measurement, and Self-Organization of Scientific Communications*. Leiden: DSWO Press, Leiden University; at http://www.upublish.com/books/leydesdorff-sci.htm.

Leydesdorff, L. (2001). Indicators of Innovation in a Knowledge-Based Economy. *Cybermetrics*, 5 (Issue 1), Paper 2, at http://www.cindoc.csic.es/cybermetrics/articles/v5i1p2.html.

Leydesdorff, L. (2003). The Mutual Information of University-Industry-Government Relations: An Indicator of the Triple Helix Dynamics. *Scientometrics*, 58 (2), 445-467.

Leydesdorff, L. (in preparation). Meaning and Translation at the Interfaces of Science: Mapping the Case of 'Stem-Cell Research,' Paper presented at the 27th Annual Meeting of the *Society for Social Studies of Science (4S)*, Altanta, GA, 15-18 October 2003.

Leydesdorff, L., & R. Zaal (1988). Co-Words and Citations. Relations between Document Sets and Environments. In L. Egghe & R. Rousseau (Eds.), *Informetrics 87/88* (pp. 105-119). Amsterdam: Elsevier.

Leydesdorff, L., & M. Meyer. (2003). The Triple Helix of University-Industry-Government Relations: Introduction to the Topical Issue. *Scientometrics*, 58 (2), 191-203.

Leydesdorff, L., & A. Scharnhorst. (2003). *Measuring the Knowledge Base: A Program of Innovation Studies.* Report to the "Förderinitiative Science Policy Studies" of the German Bundesministerium für Bildung und Forschung. Berlin: Berlin-Brandenburgische Akademie der Wissenschaften; at http://sciencepolicystudies.de/Leydesdorff&Scharnhorst.pdf.

Mansfield, E. (1991). Academic Research and Industrial Innovation. *Research Policy*, 20 (1), 1-12.

Meyer, M. (2000). Does Science Push Technology? Patents Citing Scientific Literature. *Research Policy 29*, 409-434.

Michelet, B. (1988). *L'Analyse des Associations.* Unpublished Ph.D. Thesis, Université Paris VII, Paris.

Narin, F., K. S. Hamilton, & D. Olivastro. (1997). The Increasing Link between U.S. Technology and Public Science. *Research Policy*, 26(3), 317-330.

Narin, F., & E. Noma. (1985). Is Technology Becoming Science? *Scientometrics*, 7, 369-381.

Narin, F., & D. Olivastro. (1988). Technology Indicators Based on Patents and Patent Citations. In A. F. J. van Raan (Ed.), *Handbook of Quantitative Studies of Science and Technology* (pp. 465-507). Amsterdam: Elsevier.

Narin, F., & D. Olivastro. (1992). Status Report: Linkage Beteen Technology and Science. *Research Policy*, 21, 237-249.

Nelson, R. R. (ed.). 1993. *National Innovation Systems: A comparative analysis*. New York: Oxford University Press.

Noble, D. (1977). *America by Design*. New York: Knopf.

OECD. (1988). *Biotechnology and the Changing Role of Government*. Paris: OECD.

OECD/Eurostat (1997). *Proposed Guidelines for Collecting and Interpreting Innovation Data, "Oslo Manual"*. Paris: OECD.

Ortega Priego, J. L. (2003). A Vector Space Model as a Methodological Approach to the Triple Helix Dimensionality: A Comparative Study of Biology and Biomedicine Centres of Two European National Councils from a Webometric View. *Scientometrics*, 58 (2), 429-443.

Owen-Smith J., M. Riccaboni, F. Pammolli, and W. W. Powell (2002). A Comparison of U.S. and European University-Industry Relations in the Life Sciences. *Management Science* 48(1), 24-43.

Pavitt, K. (1984). Sectoral Patterns of Technical Change: Towards a Theory and a Taxonomy. *Research Policy, 13*, 343-373.

Reddy, C., P. Wouters, & I. Aguillo. (2003). *Invisible Internet in Parts of the EU Research Area*. Deliverable of Project WISER to the European Commission. Amsterdam/Madrid: Nerdi, CINDOC.

Salton, G., & M. J. McGill. (1983). *Introduction to Modern Information Retrieval*. Auckland, etc.: McGraw-Hill.

Sampat, B. N., D. C. Mowery, A. A. Sidonis (2003). Changes in University Patent Quality after the Bayh-Dole Act: A Re-Examination. *International Journal of Industrial Organization* (forthcoming).

Sherman, C., & G. Price. (2001). *The Invisible Web*. Medford, NJ: Cyberage Books.

Wagner, C. S., & L. Leydesdorff. (2003). Mapping Global Science using International Co-Authorships: A Comparison of 1990 and 2000. In J. Guohua, R. Rousseau, W. Yishan (Eds.), *Proceedings of the 9th International Conference on Scientometrics and Informetrics* (pp. 330-340). Dalian: Dalian University of Technology Press.

White, H. D. (2003). Author Cocitation Analysis and Pearson's *r*. *Journal of the American Society for Information Science and Technology*, 54 (13), 1250-1259.

Whitley, R. D. (1984). *The Intellectual and Social Organization of the Sciences*. Oxford: Oxford University Press.

**Appendix 1**

Visual Basic 6 code for collecting the 2,827 patents in 2002 with either an inventor or assignee with a Dutch (NL) address

```
Private Sub Form_Load()

Dim strURL As String                       ' URL string
Dim intFile, N As Integer, P As Integer    ' FreeFile variable

P = 1
For N = 1 To 2827
   intFile = FreeFile()
   strURL = "http://patft.uspto.gov/netacgi/nph-_
      Parser?Sect1=PTO2&Sect2=HITOFF&u=/netahtml/search-adv.htm&r="_
      + LTrim(Str(N)) + "&f=G&l=50&d=PTXT&s1=(ISYR-2002+AND+_
      (nl.ASCO.+OR+nl.INCO.))&p=" + LTrim(Str(P)) + "&OS=isd/$/$/2002+_
      and+(acn/nl+or+icn/nl)&RS=(ISD/2002$$+AND+(ACN/nl+OR+ICN/nl))"
   Open ("C:\temp\p" + LTrim(Str(N)) + ".htm") For Output As #intFile
   Print #intFile, Inet1.OpenURL(strURL)
   Close #intFile
   If N Mod 50 = 0 Then P = P + 1
Next
End
```